\documentstyle[12pt,epsf]{article}

\newcommand{\eq}{\begin{equation}}
\newcommand{\en}{\end{equation}}
\newcommand{\eqn}{\begin{eqnarray}}
\newcommand{\enn}{\end{eqnarray}}
\newcommand{\CR}{\nonumber \\}
\newcommand{\hsp}{\hspace}
\newcommand{\rot}[3]{\left({#1}\atop{#2}\right)_{#3}}
\newcommand{\sig}{\sigma}
\newcommand{\th}{\theta}

\newcommand{\pa}{\partial}
\newcommand{\A}{\alpha}

\newcommand{\B}{\beta}
\newcommand{\ep}{\epsilon}
\newcommand{\no}{\nonumber}

\newcommand{\fr}{\frac}

\newcommand{\ra}{\rightarrow}
\newcommand{\equ}{\equiv}

\newcommand{\phA}{\phi_{\rm A}}
\newcommand{\phB}{\phi_{\rm B}}
\newcommand{\RR}{{\rm R}_2}

\newcommand{\la}{\lambda}
\newcommand{\de}{\delta}

\newcommand{\bea}{\begin{eqnarray}}
\newcommand{\ena}{\end{eqnarray}}
\newcommand{\vs}[1]{\vspace{#1 mm}}
\newcommand{\hs}[1]{\hspace{#1 mm}}

\newcommand{\p}[1]{(\ref{#1})}

\begin{document}
\begin{titlepage}

\begin{flushright}
UT-Komaba/99-11 \\
OU-HET 323 \\
hep-th/9908006
\end{flushright}

\begin{center}
{\Large \bf
Spectrum of Maxwell-Chern-Simons Theory Realized on Type IIB Brane
Configurations
}
\vs{10}

{\large
Takuhiro Kitao\footnote{e-mail address: kitao@hep1.c.u-tokyo.ac.jp}}\\
\vs{5}
{\em Institute of Physics, University of Tokyo, Komaba,
Meguro-ku, Tokyo 153-8902, Japan
}
\vs{10}

{\large 
Nobuyoshi Ohta\footnote{e-mail address: ohta@phys.sci.osaka-u.ac.jp}} \\
\vs{5}
{\em Department of Physics, Osaka University,
Toyonaka, Osaka 560-0043, Japan}\\
\vs{10}
{\bf Abstract}
\end{center}

We study the $3D$ field theory on one D3-brane stretched between $(r,s)$
and $(p,q)$5-branes. The boundary conditions are determined from the
analysis of NS5 and D5 charges of the two 5-branes. We carry out
the mode expansions for all the fields and identify the field theory as
Maxwell-Chern-Simons theory. We examine the mass spectrum to determine
the conditions for unbroken supersymmetry (SUSY) in this field theory and
compare the results with those from the brane configurations. The spectrum
is found to be invariant under the Type IIB $SL(2,{\bf Z})$-transformation.
We also discuss the theory with matters
and its S-dual configuration. The result suggests that the equivalence under
S-duality may be valid if we include all the higher modes in the theories with
matters. We also find an interesting phenomenon that SUSY enhancement happens
in the field theory after dimensional reduction from $3D$ to $2D$.

\end{titlepage}
\setcounter{footnote}{0}

\section{Introduction} 

It is believed that $4D$ $N=4$ supersymmetric Yang-Mills (SYM) theory is
self-dual under the $SL(2,{\bf Z})$ transformation. This is consistent with
the well-known fact that $4D$ $N=4$ SYM is realized as the field theory on
the D3-branes and that D3-branes are self-dual under the Type IIB
$SL(2,{\bf Z})$ transformation.

In the same way, it is suggested that $3D$ $N=4$ Abelian gauge theory with
$N_f$ flavors are equivalent in the low energy with $U(1)^{N_f -1}$ gauge
theory with bi-fundamental matters~\cite{IS}. 
This duality is also realized as the Type IIB S-duality
for the brane configuration which consists of D3-branes suspended between
NS5- or D5-branes~\cite{HW}. Such a realization of the duality in
the $3D$ field theory by the brane configuration is also discussed in
other cases~\cite{PZ,BHO2}. These are all reminiscent of
$SL(2,{\bf Z})$ invariance that $4D$ $N=4$ SYM theory has.

In theories with less supersymmetry (SUSY), it has been suggested
that the field theory realized on some brane configuration is equivalent
in the low energy to that realized on its S-dual brane
configuration~\cite{BHO3}-\cite{GK}. Except the case of $3D$ $N=4$, there
would be quantum corrections, so it is difficult to prove from the
field theoretic point of view that this kind of duality is exactly true.
It does seem to be likely, and it is quite interesting to understand
the dualities of the field theories from the string theoretic point of
view. There are some studies on the $3D$ low-energy effective theories
in the strong coupling region~\cite{BHOY}-\cite{KS}, in which such
dualities are mentioned in the field theory approach.

In this paper, we discuss the field theory realized on one D3-brane
stretched between $(r,s)$ and $(p,q)$5-branes which are rotated each
other.\footnote{The configuration for $(r,s)=(0,1)$ in our model is also
discussed in the analysis of the supergravity solutions~\cite{GGPT} and
in Matrix theory~\cite{OZ}.}  We can treat this as the theory which has
the same matter contents as the $4D$ $N=4$ Abelian gauge theory. The
boundary conditions are known for NS5-brane and D5-brane with vanishing
expectation value (VEV) of RR 0-form gauge field \cite{HW}. By mixing these
conditions with the weight of the two kinds of charges, we determine the
boundary condition for the $4D$ vector field. By considering the rotated
directions, we can also get boundary condition for the $4D$ scalars and
fermions. From these boundary conditions and equations of motion in the
bulk, we determine the modes on the D3-brane. In this way, we can
identify this field theory as $3D$ Maxwell-Chern-Simons (MCS) theory.
We also discuss the possibility to get the boundary conditions using the
near horizon limit of the supergravity solutions for the 5-branes.
But it seems difficult to obtain the correct results because
the interactions at the end points of the D3-brane with the gravity 
background are ambiguous.

We determine the masses in terms of $(p,q), (r,s)$, the Type IIB string
coupling constant $g_s$, the vacuum expectation value (VEV) $C_0$ of RR
0-form gauge field and the three angles corresponding to the
relative angles of the two 5-branes. We then find the relations among
these parameters according to the various SUSYs of this MCS theory.
We can compare these results with those in refs.~\cite{OT,KOO}, in which
the fraction of remaining SUSY of the configuration is discussed from the
analysis of the killing spinor in SUGRA. We find that our results are
the same as those of~\cite{OT,KOO}.

We also study $SL(2,{\bf Z})$ invariance of the mass spectrum.
It turns out that the CS mass is invariant under $SL(2,{\bf Z})$
transformation, which is reminiscent of $4D$ $N=4$ $SL(2,{\bf Z})$
invariance. 

We also discuss the theories with matters by adding D5-branes to the
above configuration. This theory is MCS gauge theory with flavors. It is
interesting to examine if this MCS theory with matters and its S-dual
theory are equivalent in the low energy, as the examples discussed in
ref.~\cite{HW}. The S-dual theory can be read off from the brane
configuration transformed from the original one. We find that there are
some ambiguous points in the S-dual theory; the theory is
almost $N=4$ Abelian gauge theory with flavors if we ignore the
interactions between fundamental matters and heavy modes.  The point is
that we have to be careful because a naive field theory limit changes
the structure of the moduli space when we keep fields with non-zero
masses. This means that we have to include the effects of all the
higher modes in order for these massive theories related with each other
by S-transformation to be equivalent in the low energy. 

This paper is organized as follows. In section~2, we discuss the
derivation of the boundary conditions and the
identification of the field theory using the method of mode expansion.
In section~3, we study the conditions for SUSYs from the masses
which are determined by the parameters $(p,q), (r,s), g_s$ and $C_0$ in
the previous section and compare those relations with the conditions for
SUSYs of this theory from SUGRA.
In section~4, we discuss how the spectrum changes under Type IIB
$SL(2,{\bf Z})$ transformation.
In section~5, we study theories with matter and their S-dual configuration.
In section~6, we add comments on interesting phenomena that SUSY enhancement
happens in the field theory after dimensional reduction from $3D$ to $2D$.
In section~7, we summarize our conclusions and discuss related issues and
some unsolved problems.

\section{Boundary condition and Mode expansions}

\subsection{Brane configurations in Type IIB and M-theory}

\begin{figure}
\epsfysize=5cm \centerline{\epsfbox{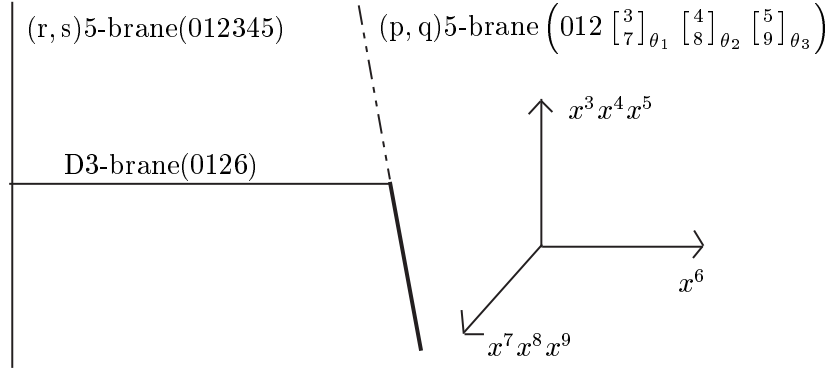}}
\caption{\small
The brane configuration that we study below.
}
\label{f0}
\end{figure}

Our brane configuration consists of two 5-branes and one D3-brane.
This D3-brane is suspended between the two 5-branes and one of its
world-volume is finite. We define the directions of the D3 world-volume as
$(x^0, x^1, x^2, x^6)$, in which the $x^6$ direction
has finite length $L$. One of the two 5-branes is $(r,s)$5-brane which has
$r$ R-R charges and $s$ NS-NS charges. Its world-volume is
$(x^0, x^1, x^2, x^3, x^4, x^5)$ and is located at $x^6=0$.
The other 5-brane is $(p,q)$5-brane whose world-volume is in the rotated
directions in $x^3$-$x^7$, $x^4$-$x^8$ and $x^5$-$x^9$ planes by three angles
$\th_1,\th_2$ and $\th_3$ relatively with the $(r,s)$5-brane. This
$(p,q)$5-brane is located at $x^6=L$ (see Fig.~\ref{f0}).

In summary, the branes we discuss below are
\begin{itemize}
\item D3-brane $(x^0,x^1,x^2,x^6 ), \hsp{2.5cm}$ at $x^{3,4,5}=x^{7,8,9}=0$,
\item $(r,s)$5-brane $(x^0,x^1,x^2,x^3,x^4,x^5), \hsp{1cm}$
at $x^6=x^{7,8,9}=0$,
\item $(p,q)$5-brane $(x^0,x^1,x^2,x^3 \cos \th_1 +
x^7 \sin \th_1, x^4 \cos \th_2 + x^8 \sin \th_2,
x^5 \cos \th_3 + x^9 \sin \th_3)$ \\
$\hspace*{3.5cm}$ at $x^6 =L$  and
$x^{k+4} \cos \th_{k-2} - x^k \sin \th_{k-2}=0, \ \ \ (k=3, 4, 5)$.
\end{itemize}

In the following, we will study how our $(r,s)$5-brane can be
represented as a 5-brane in M-theory. For $(p,q)$5-brane, we can
understand from the case of $(r,s)$5-brane by replacing $(r,s)$ with
$(p,q)$ and $\{ x^k, x^{k+4} \}$ with $\{ x^k \cos \th_{k-2}+ x^{k+4}
\sin \th_{k-2}, x^{k+4} \cos \th_{k-2} - x^k \sin \th_{k-2} \}$ for
$k=3,4,5$.

\begin{figure}[htb]
\epsfysize=5cm \centerline{\epsfbox{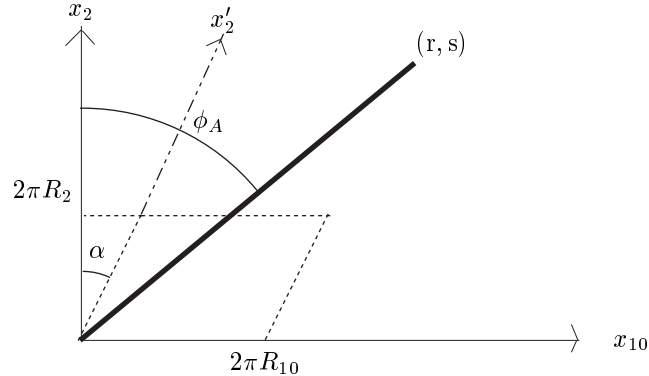}}
\caption{\small
An M5-brane wrapping a torus $T^2$ $r$ times over $x_{10}$
and $s$ times over $x_{2}'$.
}
\label{f1}
\end{figure}

Let us consider an M5-brane whose world-volume is $(x^0, x^1, x^2\cos \phA
+ x^{10} \sin \phA, x^3, x^4, x^5)$.
We can get the Type IIB 5-brane from this M5-brane by
compactifying the two directions $x^{10}$-$x^2$ on a torus and taking
T-duality~\cite{SCH}. We define the complex structure of this torus as
\bea
\tau = \fr{R_2 \tan \A }{R_{10}} +i \fr{R_2}{R_{10}},
\ena
where ${R_2}$ and ${R_{10}}$ are the radii of the compactified $x_2$ and
$x_{10}$, respectively, and $\A$ is the angle which represents the
rotation between the direction of $x_2$ and the direction which has
periodicity due to the compactification (see Fig.~\ref{f1}), denoted as
${x_2}'$. If we shrink ${R_{10}}$ and take T-duality in the direction of
${x_2}'$, we get Type IIB theory. In terms of the Type IIB string
coupling constant $g_s \equiv \frac{R_{10}}{R_2}$ and the VEV
$C_0 \equiv \tan\A/g_s$ of RR 0-form, we
can also write this complex structure as $\tau = C_0 + \fr{i}{g_s}$.

On this torus, we can depict the Type IIB $(r,s)$5-brane as the M5-brane
wrapping $x_2'$ $s$ times and $x_{10}$ $r$ times. We thus set $\phA$ as
\eq
\tan\phA
= \fr{r R_{10} + s R_2 {\tan \A}}{s R_2}
=\fr{r}{s} g_s + C_0 g_s.
\label{pha}
\en
The corresponding angle $\phi_B$ for $(p,q)$5-brane is defined by
replacing $(r,s)$ with $(p,q)$.
As a result, our brane configuration can be expressed in M-theory as
\eqn
{\rm M5}\Big[01\rot{2}{10}{\phi_A}345 \Big] -
   {\rm M2}\left(016\right) 
     - {\rm M5}\Big[01\rot{2}{10}{\phi_B}\rot{3}{7}{\th_1}
       \rot{4}{8}{\th_2}\rot{5}{9}{\th_3} \Big] \no
\enn
By analyzing the Killing spinors for the above three kinds of M-branes,
the conditions for the unbroken SUSY have been obtained in \cite{KOO}
(see also \cite{OT}, in which the authors study the unbroken SUSY of the
two rotated M5-branes). The results are summarized as follows:
\begin{itemize}
\item $\fr{1}{16}$  ${\rm SUSY}$ ($3D$ $N=1$):
$|\phi_A-\phi_B|=|\th_1 \pm \th_2 \pm \th_3 |$,
\item $\fr{1}{8}$ ${\rm SUSY}$ ($3D$ $N=2$):
$|\phi_A-\phi_B|=|\th_i|$, $|\th_j|= |\th_k|$, $(i,j,k)=(1,2,3)$,
\item $\fr{3}{16} $ ${\rm SUSY}$ ($3D$ $N=3$):
$|\phi_A-\phi_B|=|\th_1|=|\th_2|= |\th_3|$.
\end{itemize}

\subsection{Mode expansions for vector fields}

In this section, we will derive the mode expansions for the vector fields on
the D3-brane from the boundary conditions which we derive now. Let us first
consider the $4D$ field theory on the D3-brane whose world-volume is
separated by NS5-brane or D5-brane. This is the case discussed in \cite{HW}.
The action for the $4D$ gauge field with zero VEV of RR 0-form is given as
\eq
\int d^3 x dx^6 \left[ - \frac{1}{4 g_{YM_4}^2} F_{mn} F^{mn} \right],
\en
where $d^3x \equiv dx^0 dx^1 dx^2$, $m$ and $n$ run over $0, 1, 2, 6$,
and $g_{YM_4}$ is the $4D$ gauge coupling
constant in $4D$ $N=4$ SYM and is related to the string coupling constant
by $g_{YM_4}^2=2 \pi g_s$~\cite{IMSY}.

Actually the theories are confined to the region $0 \leq x^6 \leq L$ by
the two 5-branes. We are then left with surface terms from the boundaries
of $x^6$ coordinate in making the variation of the action with respect to
$A_\mu$. From the variation in the above action, we find the surface term
\eq 
\int d^3 x \delta A^\mu \frac{1}{g_{YM_4}^2} F_{\mu 6},
\label{var}
\en
at the boundaries in the $x^6$ direction.
These must vanish by themselves. Hanany and Witten
suggest that the boundary condition is $F_{\mu \nu}=0$ 
($\mu, \nu=0, 1, 2$) for D5-brane from
the gauge invariant representation of $\delta A^{\mu} =0$ and
$F_{\mu 6}=0$ for NS5-brane~\cite{HW}.

Now the question is what happens if there is non-zero VEV of RR 0-form
in the above configuration. The action for the $4D$ gauge field with non-zero
VEV of RR 0-form is given by
\eq
\int d^3 x dx^6 \Big[  - \frac{1}{4 g_{YM_4}^2} F_{mn} F^{mn} +
  \fr{C_0}{4\pi} \ep^{6 \mu \nu \la} F_{6 \mu} F_{\nu \la} \Big],
\en
where $\mu, \nu$, and $\la$ run over $0, 1, 2$. From the variation of this
action, by the same procedure as above, we are led to the boundary conditions
\eqn
\fr{1}{g_s} F_{\mu 6} + C_0 \fr{1}{2}\ep_{\mu \nu \la}F^{\nu \la}&=& 0
\ \ \ {\rm for}\ \ {\rm NS5}, \label{NS5C0} \\
\fr{1}{2}\ep_{\mu \nu \la} F^{\nu \la} &=& 0 \ \ \ {\rm for}\ \ 
{\rm D5}.
\label{D5C0}
\enn
Note that the boundary condition on NS5-brane is changed and mixed
with that of D5-brane, but the boundary condition on D5-brane is
unchanged. In fact, for non-zero value of $C_0$, NS5-brane comes to have
D5 charge by $C_0$ in addition to unit NS5 charge. This is expressed
as $[C_0,1]$ in the charge lattice, which can be seen Fig.~\ref{f1} by
counting the winding numbers in the directions of $x_{10}$ and $x_2$,
and by using 
the relation, $C_0=\fr{R_2 \tan \A}{R_{10}}$. This is a
phenomenon similar to Witten effect in $4D$ field theory~\cite{WW}. On the
other hand, D5-brane has only one D5 charge, expressed as $[1,0]$, and 
its charge is not affected
by non-zero $C_0$. So we expect that boundary condition for the 5-brane
is given by mixing these two kinds of boundary conditions at the rate of
the charges that the 5-brane has. For $(r,s)$5-brane with non zero $C_0$,
the charge can be expressed as $[sC_0 + r, s]$. We thus expect that the 
boundary condition can be obtained by replacing
$C_0$ in (\ref{NS5C0}) by $C_0 + \fr{r}{s}$:
\eq
\fr{1}{g_s} F_{\mu 6} +
   (C_0 + \fr{r}{s} ) \fr{1}{2}\ep_{\mu \nu \la} F^{\nu \la}= 0.
\label{cond}
\en
This also includes the boundary condition for D5-brane, as can be seen by
putting $(r,s)=(1,0)$. The term with the coefficient $\fr{r}{s}$ may be
understood as coming from the condensation on the $(r,s)$5-brane just
as the case of the fundamental string boundary condition on the bound
state of D-branes.
For $(p,q)$5-brane, we obtain the condition by replacing $(r,s)$ by
$(p,q)$. Therefore in our model, we get the boundary conditions for
the vector field as
\eqn
F_{\mu 6} 
  + (C_0 + \fr{r}{s}) g_s \fr{1}{2}\ep_{\mu \nu \la} 
F^{\nu \la}= 0,\ \ \ {\rm at} \ \ \ x^6=0, \CR
F_{\mu 6}
 + (C_0 +\fr{p}{q} ) g_s \fr{1}{2}\ep_{\mu \nu \la}
   F^{\nu \la}= 0,\ \ \ {\rm at} \ \ \ x^6=L.
\label{gauge_BC}
\enn

Let us discuss the spectrum under these boundary conditions.
Using (\ref{pha}) and
\bea
\tan \phi_A = (\fr{r}{s} + C_0)g_s, \hs{5}
\tan \phi_B = (\fr{p}{q} + C_0)g_s,
\ena
we can expand the $4D$ gauge fields as
\eqn
A_{\mu} &=& \sum_{n= -\infty}^{\infty}
           a_{\mu}^{(n)}  \cos [(\phi_A - \phi_B + n\pi)
                 \fr{x_6}{L} - \phi_A], \CR
A_{6} &=& \sum_{n= -\infty}^{\infty}
           a_{6}^{(n)} \sin [(\phi_A - \phi_B  + n\pi)
               \fr{x_6}{L} -\phi_A],
\label{boundary}
\enn
up to gauge transformation. Here we denote by $a_{\mu}^{(n)}$ and $a_6^{(n)}$
the fields which depend on only $3D$ coordinates $x^0,x^1,x^2$.
These expansions lead to the following representations for
the $4D$ field strengths:
\eqn
F_{\mu \nu} &=& \sum_{n= -\infty}^{\infty}
         f_{\mu \nu}^{(n)}  \cos [(\phi_A - \phi_B  + n\pi)
             \fr{x_6}{L} - \phi_A],  \CR
F_{\mu 6} &=& \sum_{n= -\infty}^{\infty}
       f_{\mu 6}^{(n)} \sin [(\phi_A - \phi_B  + n\pi) \fr{x_6}{L} -
       \phi_A ],
\enn
where $f_{\mu \nu}^{(n)} \equ \pa_{\mu} a_{\nu}^{(n)} - \pa_{\nu}
a_{\mu}^{(n)}$ and $f_{\mu 6}^{(n)} \equ \pa_{\mu} a_{6}^{(n)}
+ \fr{(\phi_A - \phi_B  + n\pi)}{L} a_{\mu}^{(n)}$. These mode expansions
satisfy the boundary conditions (\ref{gauge_BC}) if we set
\bea
f_{\mu 6}^{(n)}= \fr{1}{2} \ep_{\mu \nu \la} f^{\nu \la (n)}.
\label{fe}
\ena
In addition to the boundary conditions, these modes have to satisfy
the equations of motion in $4D$ on the D3-brane. These equations can be
expressed in terms of the corresponding modes as\footnote{The 4D
anomaly term does not contribute to the equation of motion.}
\eqn
\pa_{\mu} F^{\mu \nu} + \pa_{6} F^{6 \nu}
&=& 0 \ \ \ \ \rightarrow \ \ \ 
      \pa_{\mu} f^{\mu \nu (n)}
          -  \fr{(\phi_A - \phi_B  + n\pi)}{L} f^{\nu 6 (n)} =0,
\label{eqm1} \\
\pa_{\mu} F^{\mu 6} &=& 0  \ \ \ \  \rightarrow  \ \ \ 
      \pa_{\mu} f^{\mu 6 (n)} =0.
\label{eqm2}
\enn
Obviously eq.~(\ref{eqm2}) is satisfied due to eq.~\p{fe}. For (\ref{eqm1}),
we find
\eqn
\pa_{\mu} f^{\mu \nu (n)} &=& \pa_{\mu} (-\ep^{\mu \nu \la}
f_{\la 6}^{(n)} )= -\ep^{\mu \nu \la} \pa_{\mu}
\left[\pa_{\la} a_6^{(n)} + \fr{(\phi_A - \phi_B + n\pi)}{L}
 a_{\la}^{(n)}\right] \CR
&=& - \fr{1}{2} \ep^{\mu \nu \la}
     \fr{(\phi_A - \phi_B  + n\pi)}{L} f_{\mu \la}^{(n)}
=  \fr{(\phi_A - \phi_B  + n\pi)}{L} \left(\fr{1}{2} \ep^{\nu \mu \la}
f_{\mu \la}^{(n)}\right).
\label{CS_eq}
\enn
Thus eq.~(\ref{eqm1}) is also satisfied by our modes. In fact the boundary
conditions~(\ref{gauge_BC}) and the field equations~(\ref{eqm1}) and
(\ref{eqm2}) are obeyed if and only if we impose the condition~\p{fe}.

By eliminating $a_6^{(n)}$ from the condition~\p{fe} as in eq.~(\ref{CS_eq}),
we get an equation for the $3D$ vector field $a_{\mu}^{(n)}$:
\eq
\pa_{\nu} f^{\nu \la (n)} - \fr{(\phi_A - \phi_B  + n\pi)}{L}
   \fr{1}{2} \ep^{\mu \nu \la} f_{\mu \nu}^{ (n)}=0.
\label{cseq}
\en
The action in $3D$ which gives us eq.~\p{cseq} as the
field equation for $a_{\mu}^{ (n)}$ is
\eq
\fr{L}{2 \pi g_s} \int d^3 x  \Big[
 -\fr{1}{4} f_{\mu \nu}^{(n)} f^{\mu \nu (n)}
  -  \fr{1}{4} \fr{(\phi_A - \phi_B  + n\pi)}{L}
    \ep^{\mu \nu \la} a_{\la}^{(n)} f_{\mu \nu}^{ (n)}  \Big].
\label{csaction}
\en
This is the desired MCS action with the mass $\left|\fr{\phi_A - \phi_B
+ n\pi}{L}\right|$. To keep mass dimension one for the $3D$ gauge field,
we have added the overall coefficient $\fr{L}{2 \pi g_s}$, but in general,
the overall factor can be absorbed in the definition of $a_{\mu}$.

{}From the condition~\p{fe}, we can write the action in a form different from
(\ref{csaction}). Define
\eq
\fr{(\phi_A - \phi_B  + n\pi)}{L} {\tilde a_{\mu}}^{(n)}
  \equ \fr{(\phi_A - \phi_B  + n\pi)}{L} a_{\mu}^{(n)} +
 \pa_{\mu} a_6^{(n)},
\en
and we can rewrite \p{fe} as
\eq 
\fr{(\phi_A - \phi_B  + n\pi)}{L} {\tilde a_{\mu}}^{(n)}
 = \fr{1}{2} \ep_{\mu \nu \la} {\tilde f^{\nu \la (n)}},
\label{sdmeq}
\en
where ${\tilde f_{\mu \nu}^{(n)}}$ is the field strength for
${\tilde a_{\mu}}^{(n)}$. The action which gives us (\ref{sdmeq}) as the
equation of motion for ${\tilde a_{\mu}}^{(n)}$ is
\eq
\fr{L}{2 \pi g_s} \int d^3 x  \Big[
 \fr{1}{4} \fr{(\phi_A - \phi_B  + n\pi)}{L}
  \ep_{\mu \nu \la}   {\tilde a^{\mu (n)}}  {\tilde f^{\nu \la (n)}}
  -  \fr{1}{2} \left(\fr{\phi_A - \phi_B  + n\pi}{L}\right)^2
  {\tilde a_{\mu}}^{(n)} {\tilde a^{\mu (n)}}  \Big].
\label{sdmaction}
\en
This is the action for self-dual model (SDM) with the mass
$\left|\fr{\phi_A - \phi_B  + n\pi}{L}\right|$. This shows that MCS theory
and SDM are classically equivalent~\cite{DJ}.

\subsection{Comments on boundary conditions from supergravity solution}

Let us examine the possibility to obtain the boundary condition
for the vector field from the supergravity solution of the 5-brane.
The classical solution for Type IIB $(r,s)$5-brane can be obtained from
the solution of M5-brane following the procedure of \cite{BHO}.
The Type IIB $(r,s)$5-brane solution is\footnote{The compactification in
Fig.~\ref{f1} is facilitated by the identification $(x^2,x^{10}) \simeq
(x^2,x^{10}) + 2\pi (R_2, R_2 \tan \A ) \simeq (x^2,x^{10}) + 2\pi (0,
R_{10}),$ but this is not a simple compactification. If we define
${x^{10}}' = x^{10} - x^2 \tan\A, \;\; {x^2}' =\frac{x^2}{\cos\A}$,
the identification becomes simply $({x^2}',{x^{10}}') \simeq ({x^2}',
{x^{10}}') + 2\pi (\frac{R_2}{\cos\A}, 0) \simeq ({x^2}',{x^{10}}') +
2\pi (0, R_{10})$, to which we can apply the method in ref.~\cite{BHO} to
obtain eq.~(\ref{axi}).}
\eqn
{ds_{\rm IIB}}^2 &=& H^{1/2} (\sin^2 \phA + H  \cos^2 \phA)^{1/2}
 \left[ H^{-1} \left\{ -(dx^0)^2 + (dx^1)^2 + (dx^2)^2  \right.\right. \no \\
&& \left.\left. + (dx^3)^2 + (dx^4)^2 + (dx^5)^2 \right\}
 + (dx^6)^2 + (dx^7)^2 + (dx^8)^2
 + (dx^9)^2 \right], \no \\
\chi &=& \fr{R_2}{R_{10}  \cos \A}
  \frac{\sin \phA  \cos(\phA - \A )-
  H \sin(\phA - \A) \cos \phA }{\sin^2\phA + H  \cos^2 \phA},\no \\
e^{\phi} &=&  \fr{R_{10}}{R_{2}} 
  H^{-\fr{1}{2}}(\sin^2\phA + H  \cos^2 \phA), \no \\
&& H=1 + \fr{Q( r, s) l_s^2 }{R^2},\ \ \ \ 
  Q( r, s)= g_s \sqrt{(r+C_0 s)^2 + \fr{s^2}{g_s^2}},
\label{axi}
\enn
where $R$ is defined by $R^2=x_6^2 + x_7^2 + x_8^2 + x_9^2$
and $l_s$ is the string length. Here we denote RR 0-form gauge field and
dilaton by $\chi$ and $\phi$, respectively. Note that the asymptotic value
of the axion $\chi$ is $C_0$, as it should be. Of course, there are
backgrounds also for NSNS 2-form, RR 2-form and RR 4-form, but we will
not need their explicit forms below. For $(p,q)$5-brane, we can
get the solution from that of $(r,s)$5-brane by replacing $(r,s)$ with
$(p,q)$ and $\{ x^k, x^{k+4} \}$ with $\{ x^k \cos \th_{k-2}+ x^{k+4}
\sin \th_{k-2}, x^{k+4} \cos \th_{k-2} - x^k \sin \th_{k-2} \}$ for
$k=3,4,5$.

We can consider that the gravitational fields (graviton, dilaton,
RR-gauge fields) are weakly coupled to the fields
of $4D$ $N=4$ SYM theory. These gravity fields come from the $(p,q)$
and $(r,s)$5-branes on the ends of the D3-brane and we treat these as
the external fields and approximate them by the classical SUGRA solution.
We consider this system in the parameter
region $g_s \ll 1$, $E \ll \fr{1}{l_s}$, where $E$ is the energy scale of
the theory we are studying now, such as $3D$ SYM gauge coupling.
This is the field theory limit,
in which we can neglect the gravitational interactions and the corrections
containing the powers of ${\A}'$ and space-time is almost flat. Near
5-branes on both the ends of the D3-brane, where the distance from the 5-brane
$R$ is much less than $l_s \sqrt{Q(r, s)}$, we have to take account of the
gravitational effects as we can see from the form of the SUGRA solution.
So when we make the variation of the $4D$ $N=4$ action weakly coupled to
the gravity, it seems a good approximation to treat the equations of motion
in the bulk as those in the flat background and derive the boundary
conditions near each 5-brane in the limit $R \ll l_s \sqrt{Q(r, s)}$ of
the SUGRA solution.

Let us consider the action for the gauge field coupled with gravity:
\eq
\int d^3 x dx^6 \Big[  - \frac{\sqrt{-G}}{8 \pi}
 {\rm e}^{-\phi} F_{mn} F^{mn} +
  \fr{\chi (x^6)}{4\pi} \ep^{6 \mu \nu \la} F_{6 \mu} F_{\nu \la} \Big],
\en
where $G = \prod_{i=0,1,2,6} G_{ii} $.
Using the above supergravity solution, we see the coefficient of
$F_{mn} F^{mn}$ reduces to the constant $-\fr{1}{8\pi g_s}$.

Since the theory is confined to the region $0 \leq x^6 \leq L$ by the two
5-branes, from the variation of the above action, we find the surface terms
\eq 
\int d^3 x \delta A^\mu
  \Big[ \frac{1}{2\pi g_s} F_{\mu 6} +
   \fr{\chi(x^6)}{2\pi} \fr{1}{2}\ep_{\mu \nu \la} F^{\nu \la} \Big],
\label{var1}
\en
at $x^6=0$ and $x^6=L$. These must vanish by themselves. 
We find that our boundary conditions are obtained by requiring that
the square bracket in eq.~\p{var1} vanish at $x^6=0$ and $x^6=L$.
Thus we get
\eqn
\fr{1}{2\pi g_s} F_{\mu 6}
     +  \fr{\chi(x^6=0)}{2\pi} \fr{1}{2}\ep_{\mu \nu \la}
        F^{\nu \la}= 0,\ \ \ {\rm at} \ \ \ x^6=0, \CR
\fr{1}{2\pi g_s} F_{\mu 6}
      +  \fr{\chi(x^6=L)}{2\pi} \fr{1}{2}\ep_{\mu \nu \la}
         F^{\nu \la}= 0,\ \ \ {\rm at} \ \ \ x^6=L.
\enn
By using the Type IIB solution for $\chi (x^6)$, we can determine
$\chi (x^6)$ on the boundary as
\eq
\chi(x^6=0) \sim  - \fr{R_2}{R_{10} \cos \A}
   \frac{\sin(\phA - \A ) \cos \phA}{\cos^2 \phA}
= C_0 - \fr{{\rm tan} \phA}{g_s}
=-\fr{r}{s},
\en
where use has been made of eq.~\p{pha}.

Similarly we can get the condition at $x^6= L$ near $(p,q)$5-brane by
replacing $(r,s)$ and $\phA$ with $(p,q)$ and $\phB$, respectively.
Thus the boundary conditions in this approximation are
\eqn
F_{\mu 6} 
 - \fr{r}{s} g_s \fr{1}{2}\ep_{\mu \nu \la}
F^{\nu \la}= 0,\ \ \ {\rm at} \ \ \ x^6=0, \CR
F_{\mu 6}
 - \fr{p}{q} g_s  \fr{1}{2}\ep_{\mu \nu \la}
   F^{\nu \la}= 0,\ \ \ {\rm at} \ \ \ x^6=L.
\label{Ggauge_BC}
\enn
These conditions are different from those in (\ref{gauge_BC}).
What is the reason for the difference and which are correct? In the
present method using supergravity solutions, we have simply taken the
$(r,s)$5-brane solution~(\ref{axi}) and used its values at the boundaries.
This is the so-called probe method. However, this neglects the contribution
{}from the interaction between each 5-brane and the D3-brane at the end 
point of the D3-brane.
Also there may be additional contributions from the
difference between the equations of motion near the boundaries and in the
bulk flat background. For these reasons, we believe that the derivation
of the boundary conditions based on the 5-brane charges are more reliable.
The boundary conditions in (\ref{gauge_BC}) also give
$SL(2,{\bf Z})$-invariant spectrum, as we will discuss in section~4, and
this gives another evidence that they are the correct ones.

\subsection{Mode expansions for other fields}

Next let us consider the scalar fields. The boundary conditions for those
are the generalization of those in ref.~\cite{HW}. It is argued that the
scalar $\phi$ corresponding to the fluctuations in the direction along the
5-brane world-volume should have Neumann boundary condition $\pa_6 \phi =0$
and those in the orthogonal directions Dirichlet condition
$\pa_{\mu} \phi =0$. The conditions in our model at $x^6 =0$ are thus
\eqn
&& \pa_6 X^l =0, \CR
&& \pa_{\mu} X^{l+4} =0,  \ \ {\rm at} \ \ x^6=0,
\enn
and those at $x^6 =L$ are
\eqn
& \pa_6 [X^l  \cos \th_{l-2} + X^{l+4} \sin \th_{l-2}] =0, &  \CR
& \pa_{\mu} [-X^l \sin \th_{l-2} + X^{l+4}  \cos \th_{l-2}]=0,
&{\rm at} \ \ x^6=L,
\enn
where $l=3, 4, 5$. For the above boundary conditions, we can make
the mode expansions
\eqn
X_{k+2} = \sum_{n= -\infty}^{\infty}
  \Phi_k^{(n)} \cos [(\th_k + n\pi )\fr{x^6}{L}], \CR
X_{k+6} = \sum_{n= -\infty}^{\infty}
  \Phi_k^{(n)} \sin [(\th_k + n\pi )\fr{x^6}{L}],
\enn
where $k=1, 2, 3$, and $\{ \Phi_k^{(n)} \}$ are the real scalar fields
in $3D$ and depend only on $x^{0},x^{1},x^{2}$. Substituting these
expressions into $X^3,X^4,X^5,X^7,X^8,X^9$ in the original $4D$ action
(see the appendix for the kinetic term of a complex scalar (two real scalars)
in $4D$), we get
\eq
\fr{L}{2 \pi g_{s}}  \int d^3 x
\sum_{k=1}^{3} \sum_{n= -\infty}^{\infty} \left[
- \fr{1}{2} \pa_{\mu} \Phi_k^{(n)} \pa^{\mu} \Phi_k^{(n)}
- \fr{1}{2} \left(\fr{\th_k + n\pi }{L}\right)^2 {\Phi_k^{(n)}}^2 \right].
\label{scalaraction}
\en
This is the $3D$ massive scalar field theory with mass
$\left|\fr{ \th_k + n\pi }{L}\right|$.

Finally let us discuss the fermions. There are four $4D$ Weyl
fermions which have the origin in the $4D$ $N=4$ Abelian field theory on
the D3-brane. These four fermions transform as spinor representation under
$SO(6)$ R-transformation. Because of the relation $SO(6) \simeq SU(4)$,
they transform as fundamental representation of $SU(4)$.
The $4D$ four fermions in our model have the rotated boundary conditions
which can be obtained by transforming those for the unrotated case.

For the unrotated case, we make the variation of the fermion
kinetic terms with respect to each fermion and get
\eq
{\bar \la} {\bar \sigma^2} \delta {\la}
 -  \delta {\bar \la} {\bar \sigma^2} {\la} =0,
\en
on the boundary, where ${\bar \sigma^2} = -\sigma^2$. Here $\sigma^2$ stands
for the Pauli matrix in the $x^6$ direction.\footnote{In the appendix we
discuss the dimensional reduction from $4D$ $N=2$ SYM to $3D$ $N=4$ SYM.}
These conditions mean that ${\bar \la}= \pm \la$. If we decompose $\la$ as
$\la = \sigma^3 \fr{\la_{R} + i \la_{Im}}{\sqrt{2}}$, where $\la_R$ and
$\la_{Im}$ are real spinors, we obtain
the boundary condition $\la_{Im}=0$ for $+$ or $\la_R=0$ for $-$.
We choose the boundary condition $\la_{Im}=0$ below.

In our model, we decompose the four $4D$ Weyl spinors $\la_i$ ($i=1,2,3,4$)
into four pairs of real spinors as
\eq
\la_{l-2} = \fr{\sig^3 \left(\psi_l + i \psi_{l+4}\right)}{\sqrt{2}}, \hs{10}
\la_4 = \fr{\sig^3 \left(\psi_2 + i\psi_{10}\right)}{\sqrt{2}},
\label{Maj}
\en
where $l= 3, 4, 5$. These combinations and indices are defined
so as to correspond to the rotations of the 5-brane in the planes
$x^3$-$x^7$, $x^4$-$x^8$ and $x^5$-$x^9$ by $\th_1$, $\th_2$ and $\th_3$.
Each $\la_i$ ($i=1,2,3,4$) transforms like $\la_i \to e^{- i \th_i} \la_i$,
where $\sum_{i=1}^4 \th_i =0$ from the traceless condition for $SU(4)$.
Our above choice of the boundary conditions leads to ${\rm Im}\la_{i}=0$ at
$x^6= 0$ and ${\rm Im} {\left(e^{-i \th_i} \la_i\right)}=0$ at $x^6= L$.
We obtain
\eq
\psi_7=\psi_8=\psi_9=\psi_{10} = 0, \ \ {\rm at} \ \ x^6= 0,
\en
and
\eqn
&& -\psi_l \sin \th_{l-2} + \psi_{l+4}  \cos \th_{l-2} =0,
 \ \ (l=3, 4, 5), \CR
&& -\psi_2 \sin \th_4 + \psi_{10} \cos \th_4 =0,
 \ \ {\rm at} \ \ x^6= L.
\enn

To satisfy these boundary conditions, we expand $4D$ real fermions
in terms of $3D$ Majorana fermions as\footnote{Precisely speaking,
we also have to consider the equations of motion for the
$4D$ Weyl fermions as the case of the vector field. We use here short-cut
approach.}
\eqn
\psi_{k+2} = \sum_{n= -\infty}^{\infty}
  \Psi_k^{(n)} \cos [(\th_k + n\pi )\fr{x^6}{L}], \ \ \ \ \
\psi_{k+6} = \sum_{n= -\infty}^{\infty}
  \Psi_k^{(n)} \sin [(\th_k + n\pi )\fr{x^6}{L}], \CR
\psi_2 = \sum_{n= -\infty}^{\infty}
  \Psi_4^{(n)} \cos [(\th_4 + n\pi )\fr{x^6}{L}], \ \ \ \ \
\psi_{10} = \sum_{n= -\infty}^{\infty}
  \Psi_4^{(n)} \sin [(\th_4 + n\pi )\fr{x^6}{L}],
\enn
where $\Psi_k^{(n)}$ ($k=1, \ldots, 3$) and $\Psi_4^{(n)}$ are $3D$ Majorana
spinors. Substituting these expressions into the original $4D$ SYM on
the D3-brane, we get the $3D$ action
\eqn
&& -\fr{1}{2\pi g_s} \int d^3 x dx_6 \sum_{i=1}^4
\fr{i }{2} \Big[ {\bar \la_i} {\bar \sigma^{\mu}} \pa_{\mu} {\la_i}
 - \pa_{\mu} {\bar \la_i} {\bar \sigma^{\mu}} {\la_i}
 + {\bar \la_i} {\bar \sigma^{2}} \pa_{6} {\la_i}
 - \pa_{6} {\bar \la_i} {\bar \sigma^{2}} {\la_i} \Big] \CR
&& = \fr{L}{2\pi g_s} \int  d^3 x
   \sum_{k=1}^4  \sum_{n= -\infty}^{\infty} \fr{ i }{2} \Big[
   {\bar \Psi_k^{(n)}} \gamma^{\mu} \pa_{\mu} {\Psi_k^{(n)}}
 + \fr{\th_k + n \pi}{L}{\bar \Psi_k^{(n)}} \Psi_k^{(n)} \Big],
\enn
where $\gamma^{\mu}$ is the $3D$ gamma matrices (see the appendix for
their explicit forms and relation to Pauli matrices). This is the
$3D$ action for massive Majorana fermions with the
mass $\left|\fr{\th_k + n \pi}{L}\right|$, ($k=1, \ldots, 4$).

In summary, we have derived the $3D$ action containing vector
fields, real scalar fields and Majorana fermions. They are all massive
and have infinite sequences of higher modes.

Let us consider the theory in the limit that the higher modes decouple, in
addition to taking the field theory limit, $g_s \ll 1$, $E \ll
\fr{1}{l_s}$,
where $E$ is the energy scale of the field theory we are studying now,
such as the $3D$ SYM gauge coupling constant $g_{YM_3}^2 = 
\fr{2\pi g_s}{L}$.
Namely we consider our theory in the following parameter region:
\eqn
E \ll \fr{1}{L} \ll \fr{1}{l_s}, && g_s \ll 1, \ \ \ \
|\phi_A - \phi_B | \ll 1,   \ \ \ \ |\th_k | \ll 1; \CR
&& \fr{g_s}{L}, \ \ \fr{|\phi_A - \phi_B |}{L} \ \
{\rm and} \ \ \fr{|\th_k |}{L}\ \ \ {\rm fixed},
\label{low-energy}
\enn
where $k=1, \ldots, 4$. Suppose that we integrate out all the higher modes
with masses much larger than $E$ which is the energy scale of the field
theory in our consideration. We are then left with a theory which consists
of only $n=0$ modes:
\eqn
\fr{L}{2 \pi g_s} \int d^3 x && \hs{-5} \Big[
 -\fr{1}{4} f_{\mu \nu} f^{\mu \nu }
 - \fr{1}{4} \fr{(\phi_A - \phi_B  )}{L}
   \ep^{\mu \nu \la} a_{\la} f_{\mu \nu}
 - \fr{1}{2} \sum_{k=1}^{3} \Big( \pa_{\mu} \Phi_k \pa^{\mu} \Phi_k
 + (\fr{\th_k }{L})^2 \Phi_k^2 \Big) \CR
&& +  \fr{ i }{2} \sum_{k=1}^4  \Big(
  {\bar \Psi_k} \gamma^{\mu} \pa_{\mu} {\Psi_k}
 + \fr{\th_k}{L} {\bar \Psi_k} \Psi_k \Big) \Big],
\label{action}
\enn
where the index $(n=0)$ on each field are suppressed.
We call this limit the lowest mode limit in what follows.

\section{Condition for supersymmetry}

In this section, we discuss the number of unbroken SUSY in our
theories described by the action~(\ref{action}).
The conditions of the unbroken SUSY change according to the
various relations among the masses. On the other hand, it is known from
the analysis of the branes in M-theory which appear in our
configuration that
the four relative angles in the two M5-branes determine the unbroken SUSY
in this configuration~\cite{OT,KOO}. We can compare these conditions for
SUSY with those obtained from the analysis of the field theory on
the D3-brane.\footnote{There is no unique way to determine the signs of the
fermion mass terms. When we change the signs for the Majorana fermion in
the definition (\ref{Maj}), we can get the action with different signs
for the fermion mass terms.}

Let us study the theory (\ref{action}) in the point of the SUSY.
The combinations ($\Phi_k$,$\Psi_k$), ($k=1, 2, 3$) become $3D$ $N=1$
SUSY multiplets as we can see from the fact that they have the same mass
$\left|\fr{\th_k}{L}\right|$. But the gauge field has mass
$\left|\fr{\phi_A - \phi_B }{L}\right|$ while the fourth fermion $\Psi_4$
has the mass $\left|\fr{\th_4}{L}\right|$. These two masses are different
in general, which means that there is no unbroken SUSY.
If these two masses are the same, they constitute a $3D$ $N=1$ gauge
multiplet and we have at least $3D$ $N=1$ SUSY. Thus the condition for
$3D$ $N=1$ SUSY is
\eq
|\th_4 |=|\th_1 + \th_2 + \th_3 | = |\phi_A - \phi_B |.
\label{3D1}
\en
Note that we can express $|\phi_A - \phi_B |$ in terms of 
$(r,s)$, $(p,q)$, $g_s$ and $C_0$ as 
\eq
|{\tan} (\phA - \phB ) | =
 \left|\fr{qr-ps}{g_s( pr + sq |\tau |^2 ) + C_0 g_s (ps + qr)}\right|.
\label{2-10}
\en
The condition for SUSY (\ref{3D1}) is in agreement with the result in
the M-theory summarized in subsection 2.1 for $3D$ $N=1$, that is,
1/16 SUSY in terms of the original $10D$ $N=2$ theory.

The discussions in refs.~\cite{OT,KOO} are based on the analysis of the
system of two relatively rotated M5-branes and an M2-brane, in which they
are treated as M-branes with infinite world-volumes. As a result, the
contribution of the boundary of the M2-brane and of the intersections
between M5-branes and M2-brane could be neglected. If we do not have such
intersections in our configuration, the same SUSY is expected
when we reduce the system with the two M5-branes and one M2-brane to
that in IIA theory and convert it into the system in IIB theory by T-duality.
The VEV of RR 0-form $C_0$ has no effects on the field
theory on the D3-brane with infinite world-volume because the anomaly term
$F \wedge F$ is a total derivative and topologically trivial in $4D$ Abelian
theory. But if one direction of the D3-brane is finite by the existence of the
intersections, this anomaly term does not vanish and might cause some
discrepancy between the field theory and M-theory analyses.
The above result shows that such discrepancy does not actually appear.

We can also derive the conditions for more SUSY from the relations
among the masses of the four $3D$ $N=1$ multiplets. For example, if two of
them and the rest are the same
\eq
|\th_1 | =
\left|\phi_A - \phi_B \right|,
\ \ \ \ \ |\th_2 | = |\th_3 |,
\label{1/8}
\en
then we have $3D$ $N=2$ SUSY. In this case, we have one $N=2$ massive
vector multiplet with mass $\fr{|\th_1|}{L}$ and one $N=2$ massive
hypermultiplet with mass $\fr{|\th_2|}{L}$. 
This condition for 1/8 SUSY is again the same as the M-theory analysis
in refs.~\cite{OT,KOO}.

Moreover if all the four masses are the same
\eq
|\phi_A - \phi_B | 
= |\th_1 | =|\th_2 | = |\th_3 |,
\label{3/16}
\en
we have $3D$ $N=3$ SUSY. $3D$ $N=3$ fields consist of one
$N=2$ massive vector multiplet and one $N=2$ massive hypermultiplet
with the same mass. Here also we reproduce the
same result as that of \cite{OT,KOO}.\footnote{In $3D$ $N=3$ MCS theory,
one of the four fermion mass terms has different sign from those of
the other three masses~\cite{KLL} (also see the appendix). We can
adjust the sign of the mass term in our model by changing the sign of
the $3D$ Majorana fermion in the definition (\ref{Maj}). For example,
when we change the sign of $\psi_4$ as $\psi_4 \ra -\psi_4 $ in
(\ref{Maj}), we get the same action as~\cite{KLL} under the condition
$\phi_A - \phi_B  = \th_1 =-\th_2 = -\th_3 = \th_4 $. Only $\Psi_1$ has
different sign in the mass term from those of the other fermions.}

\section{$SL(2,{\bf Z})$ transformation and spectrum}

Let us discuss the invariance under the $SL(2,{\bf Z})$ transformation of
the type IIB brane configuration. The boundary condition~(\ref{gauge_BC})
for the gauge field has the T-invariance under
\bea
\left(\matrix{ r \cr s\cr}\right) \to
\left(\matrix{ r-s \cr s\cr}\right),\ \ \ \
\left(\matrix{ p \cr q\cr}\right) \to
\left(\matrix{ p-q \cr q\cr}\right),\ \ \ \ 
C_0 \to C_0 +1,
\ena
but the S-invariance under
\bea
\left(\matrix{ r \cr s\cr}\right) \to
\left(\matrix{ s \cr -r\cr}\right),\ \ \ \
\left(\matrix{ p \cr q\cr}\right) \to
\left(\matrix{ q \cr -p\cr}\right),\ \ \ \ 
\tau \to -\frac{1}{\tau},
\ena
is not so apparent. If there is any invariance of the spectrum
under the $SL(2,{\bf Z})$ transformation, it would be the reflection of the
$SL(2,{\bf Z})$ invariance in the $4D$ $N=4$ Abelian gauge theory realized
on one D3-brane.

Let us consider the mass spectrum in our theory. The masses
of the scalars and the fermions are invariant under the $SL(2,{\bf Z})$
transformation. It is the mass of the gauge field that may change
under this transformation. We see that this mass is determined by the
relative angle of the two M5-brane in the $x^2$-$x^{10}$ plane. The
relative angle is $SL(2,{\bf Z})$ invariant, so we can conclude 
our spectrum are invariant under $SL(2,{\bf Z})$ transformation.
In fact this can also be confirmed by direct calculation using the
transformation
\eqn
&&\left(\matrix{ r \cr s\cr}\right) \rightarrow
\left(\matrix{ a & -b \cr -c & d \cr}\right)
\left(\matrix{ r \cr s\cr}\right),\ \ \ \
\left(\matrix{ p \cr q\cr}\right) \rightarrow
\left(\matrix{ a & -b \cr -c & d \cr}\right)
\left(\matrix{ p \cr q\cr}\right),\ \ \ \  
\tau \rightarrow \fr{a \tau + b}{c \tau + d},
\enn
where $a,b,c$ and $d$ are integers and satisfy $ad-bc=1$.

\section{Comments on the theories with fundamental matters}

We have considered the theories without matter until now. Here we study
briefly how the theories with matters\footnote{This case has also been
discussed in the literature~\cite{KOO,KK,OKankoku}.} are changed, compared
with those without matters. For simplicity, we set $C_0$ to zero below.

The matters in the (anti-)fundamental representation can be
expressed by the open strings stretched between the D3-brane and
D5-branes which have the world-volume ($x^0,x^1,x^2,x^7,x^8,x^9$).
So we need to add some D5-branes in addition to the configuration we have
discussed. We consider the case with two D5-branes as a simple example.
Moreover we will discuss below theories without mass terms for the
(anti-)fundamentals and Fayet-Iliopoulos terms.
We also put $(r,s)=(0,1)$, in which the condition for SUSY is the
same as that without matter; D5-branes give no additional condition for
the SUSY in the configuration we have discussed. The corresponding
field theory is massive (MCS) Abelian gauge theory with two
fundamental matters. The bosonic part
of the action for the flavors~\cite{TA,Ka} is
\eqn
&&S_Q= - \int d^3 x \left[
 \sum_{\mu=0}^{2} \sum_{l=1}^2 \left|(\pa_{\mu} + i a_{\mu} )Q_l \right|^2
 + \sum_{l=1}^2 Q^{\dag}_l Q_l \sum_{k=1}^3 \Phi_k^2  \right. \CR
&& \hspace*{3cm} \left. + \fr{1}{2} \sum_{k=1}^3
 \left(g_{YM_3}\sum_{l=1}^2 Q^{\dag}_l \sigma^k Q_l
 +  \fr{1}{g_{YM_3}}\fr{\th_k}{L} \Phi_k\right)^2 \right], \hs{6}
Q_l \equ \rot{q_l}{{\tilde q_l}^{\dag}}{},
\label{quark}
\enn
where $l$ is the index for the two flavors, and $q_l$ and $\tilde q_l$
are the fundamental and anti-fundamental matters, respectively. Here
$g_{YM_3}$ is the 3D gauge coupling constant  
given by $g_{YM_3}^2=\fr{2\pi g_s}{L}$. 
The above action is the same as that for $3D$ $N=4$
charged hypermultiplets except the terms containing the angles.
\begin{figure}
\epsfysize=5cm \centerline{\epsfbox{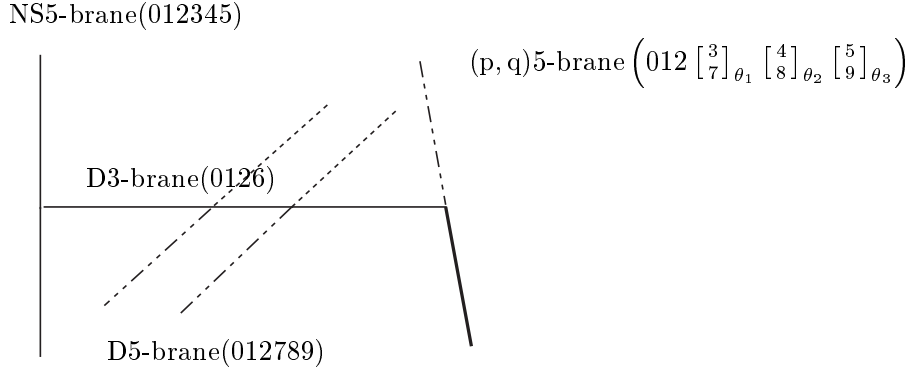}}
\caption{\small
The brane configuration corresponding to the massive
(CS) Abelian gauge theory with two fundamental matters.}
\label{f2}
\end{figure} 

Compared with ref.~\cite{HW}, there is no Coulomb branch (moduli)
parameterized by the VEV of the adjoint fields, because they are all massive
fields and cannot get VEV. In the brane configuration, this corresponds to
the fact that the D3-brane cannot move in the directions of $x^{3,4,5}$.
If the VEVs of the fundamental matters are zero, the gauge coupling goes to
the strong coupling region in the low energy and this CS
theory may become 
confining.\footnote{See \cite{IO} in which non-supersymmetric
MCS theory with matters is discussed. It is expected to be the same
in the cases of $N=1$ and $N=2$. For $N=3$, little is known
about the phase in the strong coupling.}

\begin{figure}
\epsfysize=5cm \centerline{\epsfbox{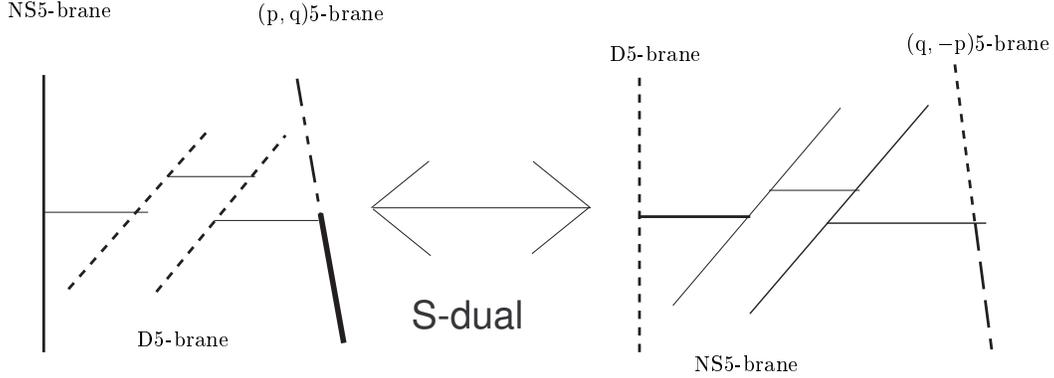}}
\caption{\small
The classical brane configuration corresponding to the Higgs phase of
the original theory and the Coulomb phase of the S-dual theory.}
\label{f3}
\end{figure}

Another phase in this theory is the Higgs, in which the
VEVs of the (anti-)fundamentals are non-zero and the gauge symmetry is
broken. This vacuum is described by $\Phi_k=0$ and
$\sum_{l=1}^2 Q^{\dag}_l \sigma^k Q_l=0$, where $k= 1, 2, 3$.
This phase can be expressed by the brane configuration in which the D3-brane
is broken into three pieces separated by the two D5-branes and the second
(middle) piece between the two D5-branes is away from the other two pieces
along the directions $x^7, x^8$ and $x^9$ of the world-volume of the
D5-branes (see Figs.~\ref{f2} and \ref{f3}). This analysis is classical,
so there may be quantum corrections.
 
Let us discuss the S-transformation of this theory.
By S-transformation, the above configuration changes into that
in which NS5-brane, $(p,q)$5-brane and two D5-branes are
replaced by D5-brane, $(q,-p)$5-brane and two NS5-branes, respectively.
The field theory on this S-dual configuration is the Abelian gauge
theory coming from the second (middle) D3-brane between the two NS5-branes.
There are also two charged flavors with respect to this Abelian
gauge symmetry. They originate from the open strings stretched between
the second D3-brane and the first (left), and between the second
D3-brane and the third (right). Their interaction terms take the same
form as those in (\ref{quark}) with $\fr{\th_k}{L}=0$, the same as the
N=4 interaction terms. In addition, the matter field
corresponding to the string stretched between the second D3-brane and
the third is also charged with respect to the massive (MCS) Abelian
gauge symmetry on the third (right) D3-brane.\footnote{The field
theory on the first D3-brane is $3D$ $N=4$ Proca field theory and there
is no gauge symmetry. This Proca field comes from the different mode
expansion from (\ref{boundary}), because in the case of $\phi_A=0$ and
$\phi_B =\fr{\pi}{2}$, we can not apply the same mode expansion as
(\ref{boundary}).} The massive Abelian fields are very heavy and their
masses are ${\cal O}(\fr{1}{L})$ because we are considering the limit in
which the rotation angles are close to $\frac{\pi}{2}$ (not
$\phi_A - \phi_B $ and $\{ \th_k \}$) in this S-dual configuration.
These large masses reduce $N=4$ to less SUSY and
the heavy MCS fields couple to the charged matter in the same way as the
interaction terms in (\ref{quark}) with large masses. The bosonic
part of the action\footnote{Here we contain only the lowest modes
of the MCS fields and ignore the other MCS modes as a simple case.}
for the flavors in the S-dual theory is then
\eqn
&& \hspace{-1cm}
S_B = - \int d^3 x \Bigg[
 \sum_{\mu=0}^{2} \sum_{l=1}^2 \left|(\pa_{\mu} + i {\tilde a_{\mu}}
 - i\de_{l2} c_{\mu}  ) B_l \right|^2
 + \sum_{l=1}^2 B^{\dag}_l B_l \sum_{k=1}^3 \left({\tilde \Phi_k}
 - \de_{l2} \Theta_k\right)^2  \CR
&& \hspace{-9mm} + \fr{1}{2} \sum_{k=1}^3
 \left({\tilde g_{YM_3}^{(1)}}
   \sum_{l=1}^2 B^{\dag}_l \sigma^k B_l \right)^2
 + \fr{1}{2} \sum_{k=1}^3
  \left( {\tilde g_{YM_3}^{(2)}} B^{\dag}_2 \sigma^k B_2 
 + \fr{\kappa_k}{{\tilde g_{YM_3}^{(2)}}} \Theta_k \right)^2
\Bigg],\; B_l \equ \rot{b_l}{{\tilde b_l}^{\dag}}{},
\label{Baction}
\enn
where ${\tilde g_{YM_3}^{(1)}}$ and ${\tilde g_{YM_3}^{(2)}}$ are the $3D$
gauge coupling constants for the gauge theories on the second 
and third D3-branes. $B_1$ and $B_2$ are the matters
corresponding to the open string stretching between the second D3-brane and
the first, and the second D3-brane and the third. Their components $b_l$
and $\tilde b_l$ are the fundamental and anti-fundamental matters,
respectively. Both the $B_1$ and $B_2$ couple to the $3D$
$N=4$ Abelian multiplet $({\tilde a_{\mu}}, {\tilde \Phi_k})$.
In addition to those, $B_2$ couples to the heavy MCS fields $(c_{\mu},
\Theta_k)$ on the third D3 with large masses.

In the lowest mode limit (\ref{low-energy})\footnote{
In the S-dual configuration, we keep the (dual) gauge coupling constants
fixed in this limit. This means $g_s \ra \infty $ and $g_{YM_3}^2
\ra \infty $ in the original configuration. In the same way, the dual
gauge coupling constants become large in the case of $g_s \ra 0 $ 
keeping $g_{YM_3}^2 $ fixed in the original configuration.
This is the original reason why we come to need all the MCS higher 
modes for the duality correspondence.}, 
these heavy CS fields should be integrated out. This yields
\eqn
&&\hspace*{-1.5cm}  S_B = - \int d^3 x \Bigg[
 \sum_{\mu=0}^{2} \sum_{l=1}^2 \left|(\pa_{\mu} + i {\tilde a_{\mu}}) B_l
 \right|^2
+ \sum_{l=1}^2 B^{\dag}_l B_l \sum_{k=1}^3 {\tilde \Phi_k}^2 \CR
&& \hspace*{4.5cm}  + \fr{1}{2} \sum_{k=1}^3
  \left({\tilde g_{YM_3}^{(1)}}
  \sum_{l=1}^2 B^{\dag}_l \sigma^k B_l \right)^2
+ \sum_{k=1}^3 {\cal O}\left(\fr{1}{\kappa_k}\right) \Bigg].
\enn
The lowest mode limit (\ref{low-energy}) is nothing but the limit $\kappa_k
\ra \infty$ in the above. If we simply take the limit and ignore the
interaction terms of ${\cal O}\left(\fr{1}{\kappa_k}\right)$, we get the
same action as that for $3D$ $N=4$ SUSY. This appears to be inconsistent with
the SUSY in the field theory realized on the original brane configuration.
On the other hand, if we keep the terms of ${\tilde a_{\mu}}$, $B_2$ and
${\tilde \Phi_k}$ suppressed by the inverse of the large masses $\kappa_k$
after the integration, the SUSY would be the same as before the integration.
Due to these two possibilities in the treatment of heavy modes, the Higgs
branch is ambiguous as follows. If we keep the higher order terms, we can
easily check that there is no Higgs branch because there is no vacuum with
non-zero VEVs for the fundamentals and $\Theta_k = {\tilde \Phi} =0$ in
the action (\ref{Baction}). This is consistent with the fact that the
original field theory does not have the (would-be corresponding) Coulomb
branch. If we simply ignore the higher order terms, the Higgs branch exists
just as in ref.~\cite{HW}. 
All of this suggests that we have to consider all the MCS higher modes 
without taking the lowest mode limit for
the equivalence of the two theories related by the S-transformation.
This is a complication that arises when we keep fields with non-zero masses.

Whatever the Higgs branch is, we can easily see that in this S-dual field
theory, there is a Coulomb branch parameterized by non-zero ${\tilde \Phi_k}$
with $B_l=0$ and $\Theta_k=0$. This phase is expressed
by the brane configuration in which the second D3-brane between the two
NS5-branes is away from the other two pieces along the directions
$x^7, x^8$ and $x^9$ of the world-volume of 
the NS5-branes (see Fig.\ref{f3}).

We have classically studied the theory with fundamental matters and their
S-dual theory. It is difficult to calculate the quantum corrections
exactly, so we cannot say definitely whether these two theories are exactly
equivalent in the low energy where the distances between the 5-branes
do not appear. However, at least at the classical level, the
Higgs branch of the original field theory seems to correspond to the Coulomb
branch of its S-dual theory, as is seen from the brane configuration. If the
two theories related by S-duality are the same theory in the low energy,
probably all the MCS higher modes must be included.

\section{Reduction to $2D$ and SUSY enhancement}

In this section, we turn to the $2D$ theory which is obtained from
(\ref{action}) by dimensional reduction in the $x^2$ direction.\footnote{
Precisely speaking, $x^2$ here is ${x^2}'$ we introduced in section~2.}
This theory is realized on the D2-brane between two NS5-D4 bound
states. This Type IIA configuration is related by T-duality to our Type IIB
brane configuration that we have discussed.

Let us focus on the case $3D$ $N=3$ SUSY. We start with the
action~(\ref{action}) under the condition (\ref{3/16}). After adjusting the
sign of the mass terms by changing the sign of the $3D$ Majorana fermion in
the definition (\ref{Maj}), we can write $3D$ $N=3$ action in the well-known
form~\cite{KLL} (see the appendix for details)
\eqn
\fr{L}{2 \pi g_s} \int d^3 x && \hs{-5} \Big[
 - \fr{1}{4} f_{\mu \nu} f^{\mu \nu }
 - \fr{\kappa}{4} \ep^{\mu \nu \la} a_{\la} f_{\mu \nu}
 - \fr{1}{2} \sum_{k=1}^{3} \Big( \pa_{\mu} \Phi_k \pa^{\mu} \Phi_k
 + \kappa^2 \Phi_k^2 \Big) \CR
&& + \fr{ i }{2} \sum_{k=2}^4  \Big(
   {\bar \Psi_k} \gamma^{\mu} \pa_{\mu} {\Psi_k}
 - \kappa {\bar \Psi_k} \Psi_k \Big)
 + \fr{ i }{2} \left({\bar \Psi_1} \gamma^{\mu} \pa_{\mu} {\Psi_1}
 + \kappa {\bar \Psi_1} \Psi_1\right) \Big],
\label{n3action}
\enn
where $\kappa$ is the CS mass. By the dimensional reduction in the $x^2$
direction, we obtain from (\ref{n3action})
\eqn
\fr{L \RR '}{g_s} \int d^2 x && \hs{-5} \Big[
\ \fr{1}{2} f_{01}^2
- \fr{1}{2} \pa_{\mu} \Phi_4 \pa^{\mu} \Phi_4
 + \kappa f_{01} \Phi_4 
 - \fr{1}{2} \sum_{k=1}^{3} \Big( \pa_{\mu} \Phi_k \pa^{\mu} \Phi_k
 + \kappa^2 \Phi_k^2 \Big) \CR
&& +  \fr{ i }{2} \sum_{k=2}^4  \Big(
   {\bar \Psi_k} \gamma^{\mu} \pa_{\mu} {\Psi_k}
 - \kappa {\bar \Psi_k} \Psi_k \Big)
 + \fr{ i }{2} \left({\bar \Psi_1} \gamma^{\mu} \pa_{\mu} {\Psi_1}
 + \kappa {\bar \Psi_1} \Psi_1\right) \Big],
\enn
where $\Phi_4$ is the dimensionally reduced $3D$ gauge field $a_2$, and we
denote the radius for the compactified direction by $\RR '\equiv
\fr{R_2}{\cos \A}$ (see Fig.~\ref{f1}).

In the above action,
the fermion masses break the $SO(4)$ invariance. However, we can eliminate
$f_{01}$ since it is an auxiliary field and also make a chiral transformation
${\Psi_1}' \equ  \gamma^{2} {\Psi_1}$ to obtain
\eqn
\fr{L \RR '}{g_s} \int d^2 x && \hs{-5} \Big[
 - \fr{1}{2} \sum_{k=1}^{4} \Big( \pa_{\mu} \Phi_k \pa^{\mu} \Phi_k
 + \kappa^2 \Phi_k^2 \Big) \CR
&& +  \fr{i}{2} \sum_{k=2}^4  \Big(
   {\bar \Psi_k} \gamma^{\mu} \pa_{\mu} {\Psi_k}
 - \kappa {\bar \Psi_k} \Psi_k \Big)
 +  \fr{i}{2} \left({\bar \Psi_1}' \gamma^{\mu} \pa_{\mu} {\Psi_1}'
 - \kappa {\bar \Psi_1}' {{\Psi_1}'}\right) \Big].
\enn
This is the sigma model action with $2D$ (4,4) SUSY, that is, 1/4 SUSY
compared with $10D$ $N=2$ theory. The SUSY is enhanced from 3/16
to 1/4 after dimensional reduction. It is crucial that as far as the
degrees of freedom are concerned, there is no difference
between $2D$ $N=(4,4)$ vector multiplet and $2D$ $N=(4,4)$ neutral
hypermultiplet except that the vector multiplet has the gauge field
without physical propagating modes; both of them consist of four real
scalars and four fermions.
The difference appears when the gauge interaction term exists. To keep
these gauge interaction terms invariant under $2D$ $N=(4,4)$ transformation,
the vector multiplet must be massless.
So we cannot construct massive gauge theory
with unbroken gauge symmetry in $2D$ $N=(4,4)$ theory.\footnote{There
is one special case in which we may expect $2D$ $N=(4,4)$ massive gauge
theory. Some kind of $3D$ gauge theory with BF term is known to have
$N=4$ SUSY. This theory does not have kinetic terms for the fermions.
If we reduce this theory to $2D$, we expect $2D$ (4,4) SUSY (see \cite{KS}
and references therein).} But in our case, the $3D$ $N=3$ massive vector
multiplet reduces to $2D$ $N=(4,4)$ massive hypermultiplet. This happens
only in Abelian gauge theory without charged matters. There is no gauge
interaction in this theory.
For non-Abelian gauge theory, $3D$ $N=3$ field theory reduces to $2D$
(3,3) non-Abelian gauge theory with some complicated interactions as well
as the mass term for the gauge field. In the same way, the $3D$ $N=3$
MCS theory with fundamental matters reduces to $2D$ (3,3) field theory
with fundamental matters. These two $N=3$ theories have gauge
interactions, so the SUSY enhancement never occurs.

{}From the above calculations of the reduction from $3D$ to $2D$,
we can say that this SUSY enhancement is accidental in $2D$ Abelian field
theory due to the existence of the chiral transformation. This is the
symmetry in $2D$ $N=(4,4)$ Abelian gauge theory, but not in $2D$ $N=(4,4)$
non-Abelian or $3D$ $N=4$ (massless) gauge theory (both for Abelian and
non-Abelian). On the other hand, the $3D$ CS term breaks one of the
four SUSY and is also not invariant under this kind of chiral
transformation. When we reduce this term from $3D$ to $2D$, the broken SUSY
transformation combined with the chiral transformation keeps the reduced
CS term as well as $2D$ (4,4) Abelian action invariant.\footnote{In the
appendix, this is shown for the Abelian field theory with the
off-shell SUSY.}
What we have found is that this transformation becomes the new SUSY
transformation and SUSY is enhanced.

On the other hand, the SUSY of the brane configuration which realizes
this $2D$ (4,4) massive theory is expected to be 3/16 \cite{OT, KOO}.
If we consider the gravity fields as well, we expect this kind of
enhancement does not occur in this sector and the SUSY remains 3/16.

\section{Conclusions and Discussions}

We have discussed the spectrum of the theory on the D3-brane separated between
$(r,s)$ and $(p,q)$5-branes. One of these 5-branes have rotated world-volumes
with the three relative angles compared with the other. We regard this
theory as the partially broken $4D$ $N=4$ theory by the boundary conditions.
Assuming that the boundary conditions for the gauge field are determined
by mixing the boundary conditions for the NS5-brane with vanishing VEV of
RR 0-form gauge field and that for D5-brane at the rate of NS5 and D5 charges,
we obtain the boundary conditions on the
intersections of the D3-brane and $(r,s)$ or $(p,q)$5-branes.
We also get the boundary conditions for the scalars and fermions by
considering the direction of the rotated 5-brane world volume.
We then find the modes which obey the boundary conditions as
well as the equations of motion for the $4D$ Abelian fields on the
D3-brane. From these modes we read off the $3D$ mass terms, which enable us
to study the conditions for SUSY and the invariance under the
Type IIB $SL(2,{\bf Z})$ transformation. We have found that the results
for SUSY are the same as those in ref.~\cite{OT, KOO}
in which the SUSY conditions are determined by the four relative angles of
the rotated two M5-branes. We have also found that this theory has
the $SL(2,{\bf Z})$ invariance. The masses we have obtained are determined
by the $SL(2,{\bf Z})$-invariant relative angles of the two 5-branes,
so the $SL(2,{\bf Z})$ invariance manifests itself even though the boundary
conditions themselves are not manifestly invariant.

We have also discussed the case with matters by adding D5-branes.
We can easily get its S-dual configuration, but identification of this
S-dual field theory involves some ambiguity. The naive lowest mode limit
in this theory is the $3D$ $N=4$ field theory, which appears to be in
conflict with the the field theory with less SUSY of the original brane
configuration. 
One possibility is that we have to include the contribution of
the MCS higher modes without taking the lowest mode limit in order that 
these two theories are equivalent in 
the low energy where the distances between the 5-branes
do not appear. That is,
it is only in special cases such as $3D$ $N=4$ \cite{HW} that these two
theories after taking the lowest mode limit are equivalent, and in general
we need all the higher modes when we also include massive adjoint fields.

We have discussed only the case of one D3-brane, that is, Abelian theory.
How about the non-Abelian theory? The boundary conditions are then non-linear,
and it is difficult, if not impossible, to carry out the mode expansions as
we have done for Abelian. As a result, it is premature to draw any definite
conclusion about the non-Abelian theories. Intuitively, it seems that the
field theory on the multiple D3-branes is $3D$ non-Abelian CS theory by the
analogy with the Abelian theories. But there are the problems to be solved
such as the quantization of the CS coefficient or the constraint of
s-rule.

In ref.~\cite{KOO}, it has been argued that for $C_0=0$ and $(r,s)=(0,1)$
in our model, there are Wilson lines whose VEVs are limited to the $p$
different values. This observation is based on the fact that in $T^2$
($x^2$-$x^{10}$) compactified M-theory point of view, there are $p$ different
positions on the M5-brane (corresponding to the $(p,q)$5-brane) to connect
the other M5-brane (corresponding to (0,1)5-brane (NS5-brane)) by the M2-brane.
These $p$ different positions in the compactified direction of $x^2$
correspond to the VEVs of the adjoint field $X_2$ in the IIA theory.
This would require some potential terms for $X_2$ in the type IIA picture.
After T-dual transformation, the VEVs of $X_2$ would turn into the Wilson
lines in the type IIB theory. In ref.~\cite{KOO}, it is suggested that
the CS coefficient may explain the existence of these $p$ vacua.
On the other hand, the field theoretic analysis in this paper is based on
the boundary conditions derived by mixing those for D5- and NS5-branes at
the rate of the charges and we believe that these conditions are the most
natural ones from the field theory viewpoint. However, this analysis does
not seem to indicate the existence of such Wilson lines and there is no
indication of the potential in T-dualized theory within our IIB boundary
conditions. It shows that there are many $3D$ CS gauge fields with heavy
CS mass terms (large CS coefficient). It seems difficult to connect the CS
coefficient to the $p$ vacua without taking this into account. There is
a single vacuum in our 3D MCS theory, but this fact seems to be in
contradiction to the above $p$ vacua originated from M-theory.
It is possible that the information about $p$ different vacua might be dropped
in the process from M-theory to Type IIB theory. If we can find boundary
conditions which reflect such information coming from M-theory, we
might construct 3D field theory that has these $p$ different vacua.
That will be a complicated theory. How to determine this theory is an
interesting subject left for future study.

\section*{\bf Acknowledgments}
T.K. would like to thank the people of Komaba Particle and Nuclear
Theory Group, K. Fukushima, T. Hotta, I. Ichinose, M. Ikehara, T. Kuroki
and T. Yoneya for useful discussions.
N.O. thanks K. Lee and P.K. Townsend for valuable comments.
It is also a pleasure to thank T. Hara for his help in drawing the figures
on this manuscript.
The work of T.K. is supported in part by the Japan Society for the
Promotion of Science under the Predoctoral Research Program (No. 10-4361).

\section*{\bf Appendix}

In this appendix, we will explain our notations in this paper and the
method of the dimensional reduction. We show how to obtain $3D$ $N=4$
action by dimensional reduction from $4D$ $N=2$ SYM as an example. We also
exhibit the $4D$ $N=2$ and $3D$ $N=4$ SUSY transformation, some of which
remain unbroken in our model. Our notations in $4D$ SYM are those of
\cite{BW} in which $4D$ $N=1$ is discussed in detail in the superspace
formalism. We discuss $SU(N_c)$ gauge theory, the Abelian case being
straightforward.

The action of the $4D$ $N=2$ SYM is
\eqn
&& \int d^4x {\cal L}_{4D} = \int d^4x \left[\fr{1}{2}
{\rm Tr} (W^{\A}W_{\A}|_{\th \th} +
 \bar W_{\dot{\A}} \bar W^{\dot{\A}}|_{{\bar \th}  {\bar \th}}) +
 2{\rm Tr} \Phi^{\dagger} e^{2V} \Phi|_{\th \th {\bar \th}
 {\bar \th}}\right] \CR
&& = \int d^4x \Big[ -\fr{1}{4} F_{mn}^{(a)}F^{mn(a)}
 - \fr{i}{2} \sum_{k=1,2}
   \Big(  \bar \la_k^{(a)} \bar \sigma^m D_m \la_k^{(a)}
 - D_m  {\bar \la_k^{(a)}} \bar \sigma^m \la_k^{(a)} \Big)
 - D_m \phi^a D^m \phi^{a \dag} \CR
&& \hspace*{2cm}    + \fr{1}{2} D_a^2 + |F^a|^2
 + \sqrt{2} (\phi^{\dag}_b \la_{2c} \la_{1a} t_{acb}
 - \phi_b \bar \la_{1a} \bar \la_{2c} t_{abc})
 + i t_{acb} D^a \phi_c \phi^{\dag}_b \Big],
\enn
where $W^{\A}$ ($V$) and $\Phi$ are the $N=1$ superfields for the vector
multiplet $(A_{m}^a,\la_{1 \A}^a, D^a)$, and the adjoint chiral
multiplet, $(\phi^a, \la_2^a, F^a )$. We take the Wess-Zumino gauge
and drop the unphysical fields. The signature is $(-1,+1,+1,+1)$.
The index for $4D$ Weyl spinor is $\A$ ($\A=1,2$), while $a$ ($a = 1,
\ldots, N_c^2-1$) is that for the non-Abelian gauge group generators $T^a$
in the fundamental representation. These generators satisfy
$[T^a,T^b]=i t_{abc} T^c$ and ${\rm Tr}T^a T^b=\fr{\delta^{ab}}{2}$.
The covariant derivatives and the field strengths are
\eqn
&&D_m \phi^a \equ \pa_m \phi^a - t_{abc} A^b_m \phi^c , \CR
&&F_{mn}^{(a)} \equ \pa_m A_n^a - \pa_n A_m^a - t_{abc}A^b_m A^c_n.
\enn
The SUSY transformations for the above theory are
\eqn
&&\delta A_m^a= -i {\bar \la_1}^a {\bar \sig_m} \xi_2
 + i {\bar \xi_2} {\bar \sig_m} \la_1^a
 + i {\bar \la_2}^a {\bar \sig_m} \xi_1
 - i {\bar \xi_1} {\bar \sig_m} \la_2^a, \CR
&&\delta \phi^a = \sqrt{2} (\xi_1 \la_1^a + \xi_2 \la_2^a), \CR
&&\delta \la_1^a = \sigma^{mn} \xi_2 F_{mn}^a + i\xi_2 D^a
 + i \sqrt{2} \sigma^{m} \bar \xi_1 D_m \phi^a
 + \sqrt{2} \xi_1 F^{a*}, \CR
&&\delta \la_2^a = - \sigma^{mn}  \xi_1 F_{mn}^a + i \xi_1 D^{a}
 + i \sqrt{2} \sigma^{m} \bar \xi_2 D_m \phi^a
 + \sqrt{2} \xi_2 F^{a}, \CR
&& \delta F^a = i \sqrt{2} \bar \xi_2 \bar \sigma^{m} D_m \la_2^a
 -2 t_{abc}\phi^c \bar \xi_2 \bar \la_1^b
 + \left(i \sqrt{2} \bar \xi_1 \bar \sigma^{m} D_m \la_1^a
 + 2 t_{abc}\phi^c \bar \xi_1 \bar \la_2^b\right)^{\ast}, \CR
&& \delta D^a = - \xi_2 \sigma^{m} D_m \bar \la_1^a
 - D_m \la_1^a \sigma^{m} \bar \xi_2
 + \xi_1 \sigma^{m} D_m \bar \la_2^a
 + D_m \la_2^a \sigma^{m} \bar \xi_1 \CR
&& \hs{20} + 2i t_{abc}\sqrt{2}(\xi_1 \la_1^c \phi^{b*}
 - \phi^b \bar \la_1^c \bar \xi_1).
\enn

Let us consider the theory dimensionally reduced to $3D$ $N=4$ from the above
$4D$ $N=2$ Lagrangian. The signature of our metric is $(-1,+1,+1)$ in $3D$ as
in $4D$ and $\ep_{012}= - \ep^{012}= +1$. We take $x^2$ as the direction of
dimensional reduction. After reduction, we rename the $x^3$ direction the new
$x^2$. The fields of $4D$ $N=2$ SYM reduce to $3D$ $N=4$ as
\eqn
&& A_2^a=C_1^a, \ \ \ \ \ \  \phi^a= \fr{-C_2^a + iC_3^a}{\sqrt{2}}, \CR
&& F^a = \fr{-D_3^a + iD_2^a}{\sqrt{2}}, \ \ \ \
 D^a= -D_1^a + t_{abc} C_2^b C_3^c, \CR
&& \la_{k \A}^a= \fr{\sigma^3 (\psi_{2k \A}^a 
 + i\psi_{2k-1 \A}^a)}{\sqrt{2}}, \ \ \ \ \
 \bar \la_{k \dot{\A}}^a= \fr{\sigma^3 (\psi_{2k \A}^a
 - i\psi_{2k-1 \A}^a)}{\sqrt{2}}, \ \ \ (k=1,2).
\enn
We define $\{ \la_2, \la_3, \la_1, -\chi \} \equiv
\{ \psi_1, \psi_2, \psi_3, \psi_4 \}$, and the $3D$ gamma matrices
by $\{ \gamma^{\mu} \equ \sigma^3 (i \sigma^2)\bar \sigma^{m} \sigma^3 \}=
(i \sigma^2,-\sigma^3,-\sigma^1)$, where $\mu=0,1,2$ and $m=0,1,3$.
We also define $\bar \psi_{\A} $ by $\bar \psi_{\A} \equ \psi_{\B}
(-i \sigma^2)_{\B \A}$ and we always keep down the indices of spinor.

The $3D$ $N=4$ Lagrangian in these notations is
\eqn
{\cal L}_{3D}
&=& -\fr{1}{4} F_{\mu \nu}^{(a)}F^{\mu \nu (a)}
 + \fr{i}{2}\sum_{k=1}^3 \bar \la_k^{(a)}
   \gamma^{\mu} D_{\mu} \la_k^{(a)}
 + \fr{i}{2} \bar \chi^{(a)}  \gamma^{\mu} D_{\mu} \chi^{(a)} \CR
&+& \fr{1}{2} \sum_{k=1}^3 \left(- D_{\mu} C^a_k D^{\mu} C^{a}_k
 + D_k^{a2}\right) 
 + \fr{it_{abc}}{2} \left(-\ep^{ijk} {\bar \la_i^a} \la_j^b C_k^c
 + 2\sum_{k=1}^3 {\bar \la_k^a} \chi^b C_k^c \right) \CR
&& \hs{20} -\fr{1}{4}  \sum_{i,j=1}^3  (C_i^a C_j^b t_{abc})^2,
\enn
where $\ep^{ijk}$ is anti-symmetric tensor and $\ep^{123}=1$.
The SUSY transformation follows from that of $4D$:
\eqn
\delta A_{\mu}^a &=& i \sum_{k=1}^3 \bar{\eta}_k \gamma_\mu \lambda_k^a
 + i \bar{\eta_0} \gamma_\mu \chi^a, \CR
\delta \lambda_k^a &=& \fr{i}{2} \gamma^{\mu} F^{\nu \rho (a)}
\epsilon_{\mu \nu \rho} \eta_k - \epsilon_{kij} (D_i^a \eta_j
 -\gamma^{\mu} D_{\mu} C_i^a \eta_j ) - t_{abc} C_k^b
 \sum_{i \neq k} C_i^c \eta_i \CR
&& + \fr{1}{2} t_{abc} \ep_{kij} C_i^b C_j^c \eta_0
 + \gamma^{\mu} D_{\mu} C_k^a \eta_0 + \eta_0 D_k^a , \CR
\delta \chi^a &=& - \gamma^{\mu} D_{\mu} \sum_{k=1}^3 C_k^a \eta_k
 - \fr{1}{2} t_{abc} \ep_{kij} C_j^b C_k^c \eta_i -\sum_{k=1}^{3}
 \eta_k D_k^a
 + \fr{1}{2} \gamma^{\mu} F^{\nu \rho (a)} \epsilon_{\mu \nu \rho} \eta_0, \CR
\delta C_k^a &=& - i \epsilon_{ijk} \bar{\eta_i} \lambda_j^a +
i \bar{\eta}_k \chi^a -i \bar{\eta}_0 \la_k^a, \CR
\delta D_k^a &=&  i \epsilon_{ijk} \bar{\eta}_i \gamma^{\mu} D_{\mu}
 \lambda_j^a
 + i \bar{\eta}_k \gamma^{\mu} D_{\mu} \chi^a
 + i t_{abc} \left(\epsilon_{ijk} \bar{\eta}_i \chi^b C_j^c 
 - \sum_{i=1}^{3} \bar{\eta}_i \la_k^b C_i^c \right)  \CR
&& \hs{-5} + it_{abc} \left(\sum_{i=1}^{3} \bar{\eta}_i \la_i^b C_k^c
 - \sum_{i=1}^{3} \bar{\eta}_k \la_i^b C_i^c\right)
 - i \bar{\eta}_0 \gamma^{\mu} D_{\mu} \la_k^a
 + i t_{abc} \bar{\eta}_0 (\epsilon_{ijk} \la_i^b C_j^c -  \chi^b C_k^c),
\enn
If we add CS term
\eqn
{\cal L}_{CS} =&& \hs{-5} \kappa \Big[-\fr{1}{2} \epsilon^{\mu\nu\rho}
   (A_\mu^a \partial_\nu A_\rho^a -
   {t_{abc}\over3}A_\mu^a A_\nu^b A_\rho^c)
 - \fr{i}{2} \left(\sum_{k=1}^{3}
   \bar{\lambda}_k^a \lambda_k^a - \bar{\chi^a}\chi^a\right) \CR
&& \hs{10}  + \sum_{k=1}^{3} C_k^a D_k^a -
   {1\over 6}t_{abc} \ep_{ijk} C_i^a C_j^b C_k^c \Big],
\enn
we have $3D$ $N=3$ SUSY~\cite{KLL}. Only three of the above $3D$ $N=4$
SUSYs survive if we add this CS term. The transformation
by $\eta_0$ is broken by this CS term. In the Abelian case, we can get
the theory with less SUSY if we change the masses for $3D$ $N=1$ multiplets,
$(A_{\mu}, \la_1)$, $(C_1,\chi)$, $(C_2,\la_3)$ and $(C_3,\la_2)$.

Finally let us make the dimensional reduction of the $3D$ $N=3$ MCS
theory to the $2D$ massive one in the $x^2$ direction. We obtain for this
$2D$ theory
\eqn
{\cal L}_{2D+CS} = && \hs{-5} \fr{1}{2}\sum_{k=1}^4
   \left(i\bar \la_k^{(a)} \gamma^{\mu} \pa_{\mu} \la_k^{(a)}
 - \pa_{\mu} C^a_k \pa^{\mu} C^{a}_k
 + D_k^{a2}\right) \CR
&& \hs{20}  +  \kappa \left(\sum_{k=1}^{4} C_k^a D_k^a
 - \sum_{i=1}^{3} \fr{i}{2} \bar{\lambda}_i^a \lambda_i^a
 + \fr{i}{2} \bar{\la_4^a} \la_4^a\right),
\label{2DCS}
\enn
where $D_4 \equ F_{01}$, $C_4 \equ A_2$ and $\la_4 \equ \chi$ and
$\mu$ runs over $0$ and $1$. Of course, this is invariant under the
original $2D$ (3,3) SUSY transformations. After we replace $\la_k$ $(k=1,
\ldots, 4)$ by $\gamma^2 \la_k$ in the $\eta_0$ transformation reduced
{}from $3D$ to $2D$, we get
\eqn
&&\delta C_i^a = i \left(\overline{\gamma^2 \eta_0}\right) \la_i^a , \ \ \ 
\delta \la_i^a = \left[D_i^a - \gamma^{\mu} \pa_{\mu} C_i\right]
  \left(\gamma^2 \eta_0\right), \ \ \
\delta D_i^a = -i \left(\overline{\gamma^2 \eta_0}\right) \gamma^{\mu}
  \pa_{\mu} \la_i^a , \CR
&&\delta C_4^a = i \bar{\eta_0} \la_i^a , \ \ \ 
\delta \la_4^a = - \left[D_4^a + \gamma^{\mu} \pa_{\mu} C_4\right]
  \eta_0, \ \ \
\delta D_4^a = i \bar{\eta_0} \gamma^{\mu} \pa_{\mu} \la_4^a ,
\enn
where $i = 1, 2, 3$.
Under this transformation, we find that the $2D$ massive
Lagrangian~(\ref{2DCS}) is invariant. This is the new SUSY transformation
besides the original $2D$ (3,3) SUSY and the $2D$ massive theory described
by (\ref{2DCS}) has $2D$ (4,4) SUSY.

\newcommand{\NP}[1]{Nucl.\ Phys.\ {\bf #1}}
\newcommand{\PL}[1]{Phys.\ Lett.\ {\bf #1}}
\newcommand{\PR}[1]{Phys.\ Rev.\ {\bf #1}}
\newcommand{\PRL}[1]{Phys.\ Rev.\ Lett.\ {\bf #1}}
\newcommand{\PRE}[1]{Phys.\ Rep.\ {\bf #1}}
\newcommand{\PTP}[1]{Prog.\ Theor.\ Phys.\ {\bf #1}}
\newcommand{\PTPS}[1]{Prog.\ Theor.\ Phys.\ Suppl.\ {\bf #1}}
\newcommand{\MPL}[1]{Mod.\ Phys.\ Lett.\ {\bf #1}}
\newcommand{\JHEP}[1]{JHEP\ {\bf #1}}
\newcommand{\IJMP}[1]{Int.\ Jour.\ Mod.\ Phys.\ {\bf #1}}
\newcommand{\JP}[1]{Jour.\ Phys.\ {\bf #1}}

\end{document}